\documentclass[letterpaper,compsoc,twoside]{IEEEtran}
\usepackage{fixltx2e} % LaTeX patches, \textsubscript
\usepackage{cmap} % fix search and cut-and-paste in Acrobat
\usepackage{ifthen}
\usepackage[T1]{fontenc}
\usepackage[utf8]{inputenc}
\usepackage{amsmath}

\usepackage[font={small,it},labelfont=bf]{caption}
\usepackage{float}

\pdfoutput=1
\usepackage{scipy}
\makeatletter
\def\PY@reset{\let\PY@it=\relax \let\PY@bf=\relax%
    \let\PY@ul=\relax \let\PY@tc=\relax%
    \let\PY@bc=\relax \let\PY@ff=\relax}
\def\PY@tok#1{\csname PY@tok@#1\endcsname}
\def\PY@toks#1+{\ifx\relax#1\empty\else%
    \PY@tok{#1}\expandafter\PY@toks\fi}
\def\PY@do#1{\PY@bc{\PY@tc{\PY@ul{%
    \PY@it{\PY@bf{\PY@ff{#1}}}}}}}
\def\PY#1#2{\PY@reset\PY@toks#1+\relax+\PY@do{#2}}

\expandafter\def\csname PY@tok@gd\endcsname{\def\PY@tc##1{\textcolor[rgb]{0.63,0.00,0.00}{##1}}}
\expandafter\def\csname PY@tok@gu\endcsname{\let\PY@bf=\textbf\def\PY@tc##1{\textcolor[rgb]{0.50,0.00,0.50}{##1}}}
\expandafter\def\csname PY@tok@gt\endcsname{\def\PY@tc##1{\textcolor[rgb]{0.00,0.27,0.87}{##1}}}
\expandafter\def\csname PY@tok@gs\endcsname{\let\PY@bf=\textbf}
\expandafter\def\csname PY@tok@gr\endcsname{\def\PY@tc##1{\textcolor[rgb]{1.00,0.00,0.00}{##1}}}
\expandafter\def\csname PY@tok@cm\endcsname{\let\PY@it=\textit\def\PY@tc##1{\textcolor[rgb]{0.25,0.50,0.56}{##1}}}
\expandafter\def\csname PY@tok@vg\endcsname{\def\PY@tc##1{\textcolor[rgb]{0.73,0.38,0.84}{##1}}}
\expandafter\def\csname PY@tok@m\endcsname{\def\PY@tc##1{\textcolor[rgb]{0.13,0.50,0.31}{##1}}}
\expandafter\def\csname PY@tok@mh\endcsname{\def\PY@tc##1{\textcolor[rgb]{0.13,0.50,0.31}{##1}}}
\expandafter\def\csname PY@tok@cs\endcsname{\def\PY@tc##1{\textcolor[rgb]{0.25,0.50,0.56}{##1}}\def\PY@bc##1{\setlength{\fboxsep}{0pt}\colorbox[rgb]{1.00,0.94,0.94}{\strut ##1}}}
\expandafter\def\csname PY@tok@ge\endcsname{\let\PY@it=\textit}
\expandafter\def\csname PY@tok@vc\endcsname{\def\PY@tc##1{\textcolor[rgb]{0.73,0.38,0.84}{##1}}}
\expandafter\def\csname PY@tok@il\endcsname{\def\PY@tc##1{\textcolor[rgb]{0.13,0.50,0.31}{##1}}}
\expandafter\def\csname PY@tok@go\endcsname{\def\PY@tc##1{\textcolor[rgb]{0.20,0.20,0.20}{##1}}}
\expandafter\def\csname PY@tok@cp\endcsname{\def\PY@tc##1{\textcolor[rgb]{0.00,0.44,0.13}{##1}}}
\expandafter\def\csname PY@tok@gi\endcsname{\def\PY@tc##1{\textcolor[rgb]{0.00,0.63,0.00}{##1}}}
\expandafter\def\csname PY@tok@gh\endcsname{\let\PY@bf=\textbf\def\PY@tc##1{\textcolor[rgb]{0.00,0.00,0.50}{##1}}}
\expandafter\def\csname PY@tok@ni\endcsname{\let\PY@bf=\textbf\def\PY@tc##1{\textcolor[rgb]{0.84,0.33,0.22}{##1}}}
\expandafter\def\csname PY@tok@nl\endcsname{\let\PY@bf=\textbf\def\PY@tc##1{\textcolor[rgb]{0.00,0.13,0.44}{##1}}}
\expandafter\def\csname PY@tok@nn\endcsname{\let\PY@bf=\textbf\def\PY@tc##1{\textcolor[rgb]{0.05,0.52,0.71}{##1}}}
\expandafter\def\csname PY@tok@no\endcsname{\def\PY@tc##1{\textcolor[rgb]{0.38,0.68,0.84}{##1}}}
\expandafter\def\csname PY@tok@na\endcsname{\def\PY@tc##1{\textcolor[rgb]{0.25,0.44,0.63}{##1}}}
\expandafter\def\csname PY@tok@nb\endcsname{\def\PY@tc##1{\textcolor[rgb]{0.00,0.44,0.13}{##1}}}
\expandafter\def\csname PY@tok@nc\endcsname{\let\PY@bf=\textbf\def\PY@tc##1{\textcolor[rgb]{0.05,0.52,0.71}{##1}}}
\expandafter\def\csname PY@tok@nd\endcsname{\let\PY@bf=\textbf\def\PY@tc##1{\textcolor[rgb]{0.33,0.33,0.33}{##1}}}
\expandafter\def\csname PY@tok@ne\endcsname{\def\PY@tc##1{\textcolor[rgb]{0.00,0.44,0.13}{##1}}}
\expandafter\def\csname PY@tok@nf\endcsname{\def\PY@tc##1{\textcolor[rgb]{0.02,0.16,0.49}{##1}}}
\expandafter\def\csname PY@tok@si\endcsname{\let\PY@it=\textit\def\PY@tc##1{\textcolor[rgb]{0.44,0.63,0.82}{##1}}}
\expandafter\def\csname PY@tok@s2\endcsname{\def\PY@tc##1{\textcolor[rgb]{0.25,0.44,0.63}{##1}}}
\expandafter\def\csname PY@tok@vi\endcsname{\def\PY@tc##1{\textcolor[rgb]{0.73,0.38,0.84}{##1}}}
\expandafter\def\csname PY@tok@nt\endcsname{\let\PY@bf=\textbf\def\PY@tc##1{\textcolor[rgb]{0.02,0.16,0.45}{##1}}}
\expandafter\def\csname PY@tok@nv\endcsname{\def\PY@tc##1{\textcolor[rgb]{0.73,0.38,0.84}{##1}}}
\expandafter\def\csname PY@tok@s1\endcsname{\def\PY@tc##1{\textcolor[rgb]{0.25,0.44,0.63}{##1}}}
\expandafter\def\csname PY@tok@gp\endcsname{\let\PY@bf=\textbf\def\PY@tc##1{\textcolor[rgb]{0.78,0.36,0.04}{##1}}}
\expandafter\def\csname PY@tok@sh\endcsname{\def\PY@tc##1{\textcolor[rgb]{0.25,0.44,0.63}{##1}}}
\expandafter\def\csname PY@tok@ow\endcsname{\let\PY@bf=\textbf\def\PY@tc##1{\textcolor[rgb]{0.00,0.44,0.13}{##1}}}
\expandafter\def\csname PY@tok@sx\endcsname{\def\PY@tc##1{\textcolor[rgb]{0.78,0.36,0.04}{##1}}}
\expandafter\def\csname PY@tok@bp\endcsname{\def\PY@tc##1{\textcolor[rgb]{0.00,0.44,0.13}{##1}}}
\expandafter\def\csname PY@tok@c1\endcsname{\let\PY@it=\textit\def\PY@tc##1{\textcolor[rgb]{0.25,0.50,0.56}{##1}}}
\expandafter\def\csname PY@tok@kc\endcsname{\let\PY@bf=\textbf\def\PY@tc##1{\textcolor[rgb]{0.00,0.44,0.13}{##1}}}
\expandafter\def\csname PY@tok@c\endcsname{\let\PY@it=\textit\def\PY@tc##1{\textcolor[rgb]{0.25,0.50,0.56}{##1}}}
\expandafter\def\csname PY@tok@mf\endcsname{\def\PY@tc##1{\textcolor[rgb]{0.13,0.50,0.31}{##1}}}
\expandafter\def\csname PY@tok@err\endcsname{\def\PY@bc##1{\setlength{\fboxsep}{0pt}\fcolorbox[rgb]{1.00,0.00,0.00}{1,1,1}{\strut ##1}}}
\expandafter\def\csname PY@tok@kd\endcsname{\let\PY@bf=\textbf\def\PY@tc##1{\textcolor[rgb]{0.00,0.44,0.13}{##1}}}
\expandafter\def\csname PY@tok@ss\endcsname{\def\PY@tc##1{\textcolor[rgb]{0.32,0.47,0.09}{##1}}}
\expandafter\def\csname PY@tok@sr\endcsname{\def\PY@tc##1{\textcolor[rgb]{0.14,0.33,0.53}{##1}}}
\expandafter\def\csname PY@tok@mo\endcsname{\def\PY@tc##1{\textcolor[rgb]{0.13,0.50,0.31}{##1}}}
\expandafter\def\csname PY@tok@mi\endcsname{\def\PY@tc##1{\textcolor[rgb]{0.13,0.50,0.31}{##1}}}
\expandafter\def\csname PY@tok@kn\endcsname{\let\PY@bf=\textbf\def\PY@tc##1{\textcolor[rgb]{0.00,0.44,0.13}{##1}}}
\expandafter\def\csname PY@tok@o\endcsname{\def\PY@tc##1{\textcolor[rgb]{0.40,0.40,0.40}{##1}}}
\expandafter\def\csname PY@tok@kr\endcsname{\let\PY@bf=\textbf\def\PY@tc##1{\textcolor[rgb]{0.00,0.44,0.13}{##1}}}
\expandafter\def\csname PY@tok@s\endcsname{\def\PY@tc##1{\textcolor[rgb]{0.25,0.44,0.63}{##1}}}
\expandafter\def\csname PY@tok@kp\endcsname{\def\PY@tc##1{\textcolor[rgb]{0.00,0.44,0.13}{##1}}}
\expandafter\def\csname PY@tok@w\endcsname{\def\PY@tc##1{\textcolor[rgb]{0.73,0.73,0.73}{##1}}}
\expandafter\def\csname PY@tok@kt\endcsname{\def\PY@tc##1{\textcolor[rgb]{0.56,0.13,0.00}{##1}}}
\expandafter\def\csname PY@tok@sc\endcsname{\def\PY@tc##1{\textcolor[rgb]{0.25,0.44,0.63}{##1}}}
\expandafter\def\csname PY@tok@sb\endcsname{\def\PY@tc##1{\textcolor[rgb]{0.25,0.44,0.63}{##1}}}
\expandafter\def\csname PY@tok@k\endcsname{\let\PY@bf=\textbf\def\PY@tc##1{\textcolor[rgb]{0.00,0.44,0.13}{##1}}}
\expandafter\def\csname PY@tok@se\endcsname{\let\PY@bf=\textbf\def\PY@tc##1{\textcolor[rgb]{0.25,0.44,0.63}{##1}}}
\expandafter\def\csname PY@tok@sd\endcsname{\let\PY@it=\textit\def\PY@tc##1{\textcolor[rgb]{0.25,0.44,0.63}{##1}}}

\def\PYZus{\char`\_}

\def\PYZsq{\char`\'}

\makeatother

\providecommand*{\DUrole}[2]{%
  \ifcsname DUrole#1\endcsname%
    \csname DUrole#1\endcsname{#2}%
  \else% backwards compatibility: try \docutilsrole#1{#2}
    \ifcsname docutilsrole#1\endcsname%
      \csname docutilsrole#1\endcsname{#2}%
    \else%
      #2%
    \fi%
  \fi%
}

\ifthenelse{\isundefined{\hypersetup}}{
  \usepackage[colorlinks=true,linkcolor=blue,urlcolor=blue]{hyperref}
  \urlstyle{same} % normal text font (alternatives: tt, rm, sf)
}{}

\begin{document}
\newcounter{footnotecounter}\title{High-Content Digital Microscopy with Python}\author{Fabrice Salvaire$^{\setcounter{footnotecounter}{1}\fnsymbol{footnotecounter}\setcounter{footnotecounter}{2}\fnsymbol{footnotecounter}}$%
          \setcounter{footnotecounter}{1}\thanks{\fnsymbol{footnotecounter} %
          Corresponding author: \protect\href{mailto:fabrice.salvaire@orange.fr}{fabrice.salvaire@orange.fr}}\setcounter{footnotecounter}{2}\thanks{\fnsymbol{footnotecounter} Genomic Vision SA}\thanks{%

          \noindent%
          Copyright\,\copyright\,2014 Fabrice Salvaire. This is an open-access article distributed under the terms of the Creative Commons Attribution License, which permits unrestricted use, distribution, and reproduction in any medium, provided the original author and source are credited. http://creativecommons.org/licenses/by/3.0/%
        }}\maketitle
          \renewcommand{\leftmark}{PROC. OF THE 6th EUR. CONF. ON PYTHON IN SCIENCE (EUROSCIPY 2013)}
          \renewcommand{\rightmark}{HIGH-CONTENT DIGITAL MICROSCOPY WITH PYTHON}

\setcounter{page}{29}
\newcommand*{\docutilsroleref}{\ref}
\newcommand*{\docutilsrolelabel}{\label}
\AtEndDocument{\cleardoublepage}

% -------------------------------------------------------------------------------------------------
\begin{abstract}High-Content Digital Microscopy enhances user comfort, data storage and
analysis throughput, paving the way to new researches and medical
diagnostics. A digital microscopy platform aims at capturing an
image of a cover slip, at storing information on a file server and a database,
at visualising the image and analysing its content. We will discuss how the
Python ecosystem can provide such software framework efficiently. Moreover
this paper will give an illustration of the data chunking approach to
manage the huge amount of data.\end{abstract}

\section{Introduction%
  \label{introduction}%
}

Since early times optical microscopy plays an important role in biology
research and medical diagnostic. Nowadays digital microscopy is a natural
evolution of the technology that provides many enhancements on user comfort,
data storage and analysis throughput. First, in comparison to binocular
microscopy where the low light emission intensity of the specimens causes
sever stress to the eyes, the digital microscopy monitor display offers
greater comfort to the users. Second, the digitization of the output allows to
freeze and store information for short to long term storage, to compress the
data, and to easily duplicate it, to protect its integrity (by checksum) and
its confidentiality (by cryptography). On the other hand, optical microscopy
implies conservation of the specimens themselves at low temperature and in the
dark. Last, the automation of a high content application provides a
considerable scale-up of the data processing throughput, thus paving the way
to new researches and medical diagnostics.

We will discuss in this paper how the Python ecosystem can provide efficiently
a software framework for the digital microscopy. Our discussion will first
present the data acquisition method, then we will describe the data storage
and finally the image viewer.

\section{Data Acquisition%
  \label{data-acquisition}%
}

The first challenge of high-content digital microscopy is the quantity of data. Let us
first evaluate how large the data is, and enlighten our reader of the reasons
of such quantity of data.
To reach the required resolution to
see the details of a specimen, optical microscopes use objectives magnifying up to the
diffraction limit which is about $100\times$. Nowadays the pixel size for a CCD and sCMOS
camera is about $6.5\,\text{um}$, thus we reach a resolution
of $162.5\,\text{nm}$ at a magnification of $40\times$.
Now consider a specimen put on a cover
slip: a glass square surface of 18 mm wide (we will later relate the support and the
specimen by the more generic word \emph{slide}, which corresponds to a larger glass surface). Consequently
to cover this surface at such magnification we have to acquire an area larger than $100\,000$
px wide, thus of the order of 10 billion of pixels. This is roughly 300 times larger than the actual
largest professional digital camera ($36\,\text{Mpx}$). In light of foregoing digital
microscopy are big data similar to spatial images and involve a software framework similar to the
well-known Google Map.

For scientific application, we preferably use monochrome camera so as to avoid the interpolation of
a Bayer mosaic. Instead, to capture the entire colour spectrum at the same time, colours are captured
sequentially by placing a filter of the colour's corresponding wave length transmission in front of
the camera. These shots are called \emph{colour fields of view} here. Figure \DUrole{ref}{epifluorescence-microscope}
shows the schematic of an application of this acquisition method called an
epifluorescence microscope.\begin{figure}[bht]\noindent\makebox[\columnwidth][c]{\includegraphics[scale=0.50]{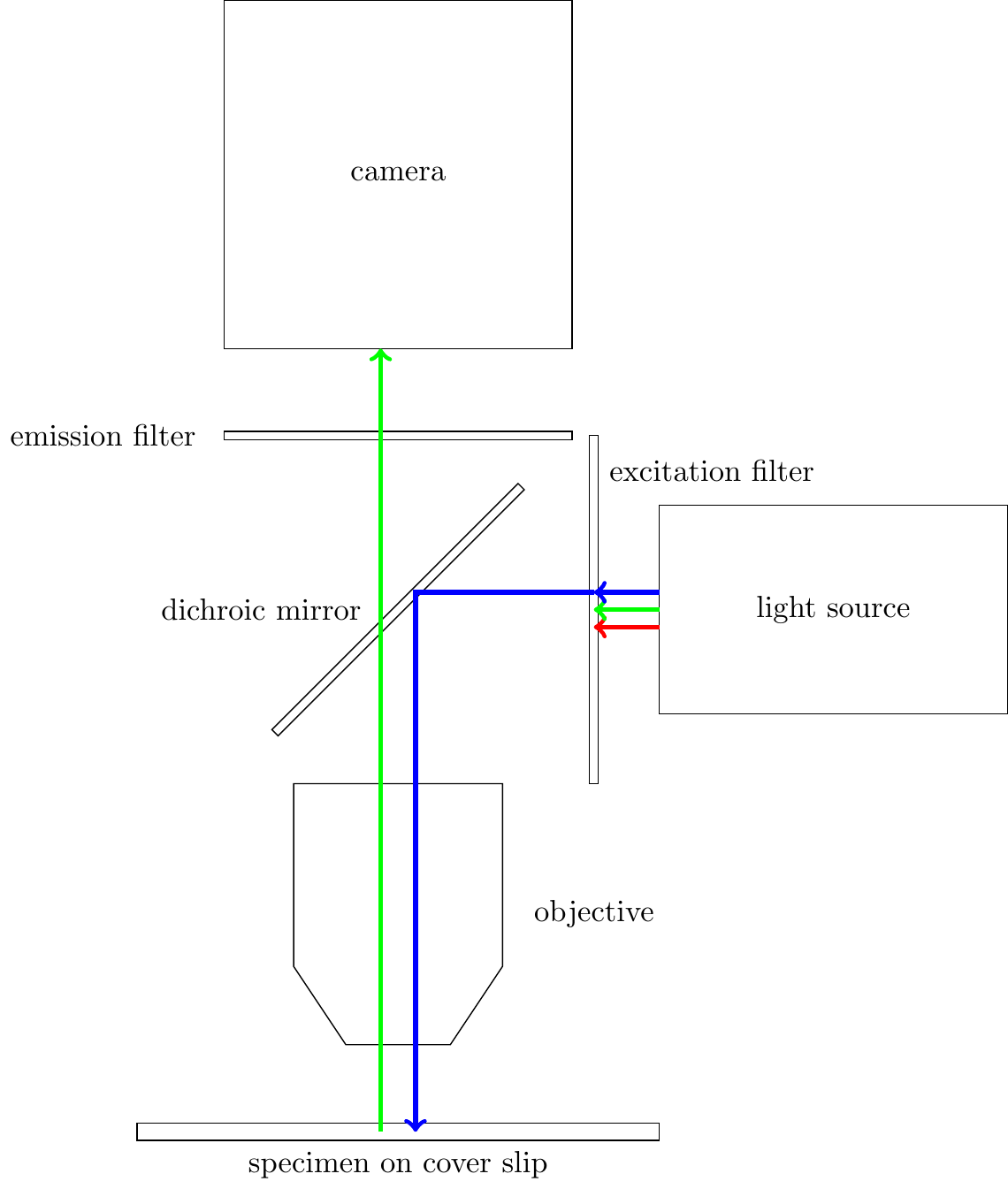}}
\caption{Schematic of an epifluorescence microscope. Specimens are labelled with fluorescent
molecules so called fluorophores. In this example we are capturing an image for a fluorophore
having an excitation wave length in the blue and an emission wave length in the green. The
filters are used to restrict the excitation and filter the
emission, respectively. \DUrole{label}{epifluorescence-microscope}}
\end{figure}

A camera like the Andor Neo sCMOS features a sensor of resolution $2560 \times 2160\,\text{px}$
and a surface of $416 \times 351\,\text{um}$. Thus to cover
the whole specimen surface we have to capture a mosaic of fields of view of size $43 \times
51$ (2193 tiles) using an automated stagger. We will also refer to the fields of view as
\emph{tiles} or \emph{images} according to the context.
% on the mosaic which depends of the step positioning error

To observe the specimen in several colours, two strategies can be used to
acquire the mosaic: one is to acquire a mosaic per colour and the other is to acquire several colours per
field of view. Both methods have advantages and drawbacks. One of the differences is the
uncertainty that occurs on the registration of the colour fields of view. When capturing several
colours per field of view at the same staging position, the relative positioning error is due to the
optical path. When capturing a mosaic per colour, the error is also due to the
reproducibility of the stagger. On the other hand the accuracy of the tile positions is always due
to the stagger precision. So as to perform a field of view
registration without black zone in the reconstructed image, we drive the stagger with a sufficient
overlapping zone on both directions. Another irregularity on the mosaic is due to the
camera alignment error according to the stagger axes that draw a sheared mosaic pattern
(figure \DUrole{ref}{sheared-mosaic}). The shearing doesn't have any serious effect on the reconstructed image
since it only displaces systematically the fields of view in the mosaic frame.\begin{figure}[bht]\noindent\makebox[\columnwidth][c]{\includegraphics[scale=0.42]{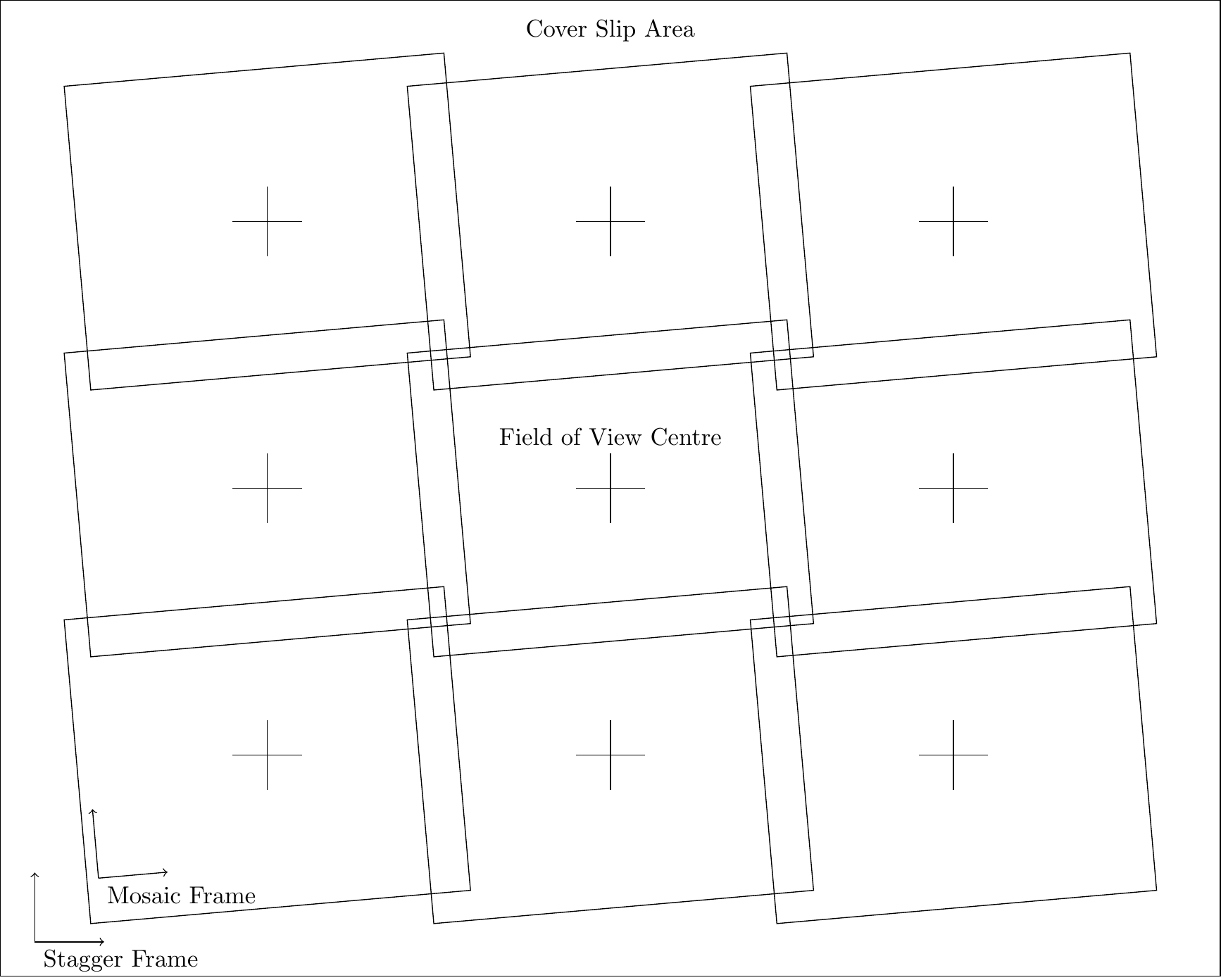}}
\caption{Field of View Mosaic showing a sheared effect due to the camera misalignment. The tiles are
rotated in the stagger frame but not in the mosaic frame. \DUrole{label}{sheared-mosaic}}
\end{figure}

All these uncertainties can be studied using fluorescent beads with an appropriate density on the
cover slip and an image registration algorithm.

The third dimension of a specimen can be observed using the vertical focus axis of the microscope
so as to perform a so called \emph{z-stack} of images that enlarge the depth of field virtually and thus
improve the focus accuracy.

The Neo camera features a standard amplifier-DAC stage with a 12-bit resolution and
another stage with a combination of two amplifier-DACs to achieve a 16-bit resolution for high
dynamic image. Thus image pixels must be encoded using an unsigned 16-bit integer data type. It
means a colour field of view weights $10.5\,\text{MB}$ and our mosaic weights
$23\,\text{GB}$ per colour.

Depending of the intensity dynamic of the specimen and the zero-padding arising from the DAC, most
of the pixels can have many zeros on the most significant bits. Therefore, the amount of
data can be efficiently reduced using a lossless compression algorithm in conjunction with a bit
shuffling to group the zeros together and form long zero sequences in the byte stream.

\section{Virtual Slide Format and Storage%
  \label{virtual-slide-format-and-storage}%
}

We can now define the data structure of an acquisition so called later a \emph{virtual slide}.  A virtual
slide is made of a mosaic of fields of view and a set of attributes that constitute the so called
\emph{slide header}. Examples of attributes are a slide identifier, a date of acquisition or a type of assay.

The mosaic is a set of colour fields of view made of a mosaic index $(r,c)$, a stagger
position $(x,y,z)$, a colour index $w$ and an image array of unsigned 16-bit integers.

From this mosaic of field of views, we can imagine to reconstruct the slide image once and for all
and produce a giant image, where we could use for this purpose the BigTIFF \cite{BigTIFF} extension to
the TIFF format. But if we want to keep raw data without information loss we have to imagine a way
to store the original fields of view and process them on-line. This case is particularly important
when the registration matters for the interpretation of the reconstructed image.

The HDF5 \cite{HDF5} library and its h5py \cite{h5py} Python binding are perfectly suited for this
purpose. The content of an HDF5 file is self-defined and the library is open source which guaranties a
long term access to the data. The structure of an HDF5 file is similar to a file system having
folder objects so called \emph{groups} and N-dimensional array objects so called \emph{dataset} that
corresponds here to files. Each of these objects can have attached attributes.  This virtual file
system provides the same flexibility than a real file system similar to a UNIX loop device. Figure
\DUrole{ref}{hdf5-file-system} shows an example.\begin{figure}[bht]\noindent\makebox[\columnwidth][c]{\includegraphics[scale=0.60]{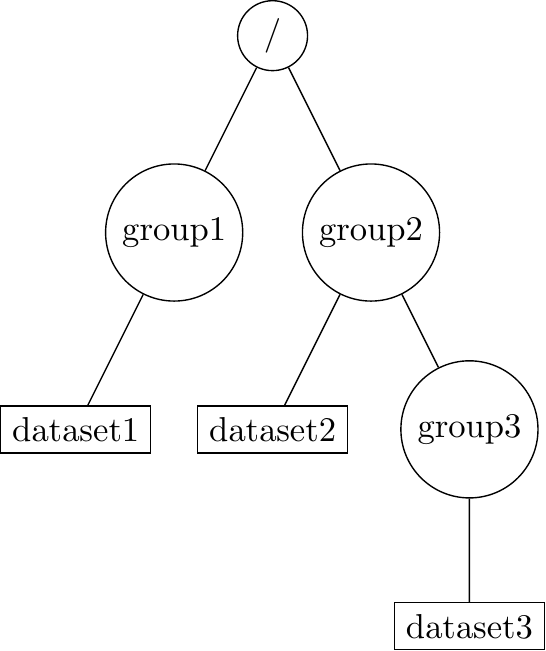}}
\caption{HDF5 Virtual File System. Attributes can be attached to each node. \DUrole{label}{hdf5-file-system}}
\end{figure}

The h5py module provides a Pythonic API and map Numpy \cite{Numpy} arrays to datasets and reciprocally.
The Numpy library is well appropriate to store images in memory since it maps efficiently a C linear
array data structure on Python. The following code snippet gives an overview of its usage:\begin{Verbatim}[commandchars=\\\{\},fontsize=\footnotesize]
\PY{k+kn}{import} \PY{n+nn}{numpy} \PY{k+kn}{as} \PY{n+nn}{np}
\PY{k+kn}{import} \PY{n+nn}{h5py}
\PY{n}{slide\PYZus{}file} \PY{o}{=} \PY{n}{h5py}\PY{o}{.}\PY{n}{File}\PY{p}{(}\PY{l+s}{\PYZsq{}}\PY{l+s}{slide.hdf5}\PY{l+s}{\PYZsq{}}\PY{p}{,} \PY{l+s}{\PYZsq{}}\PY{l+s}{w}\PY{l+s}{\PYZsq{}}\PY{p}{)}
\PY{n}{slide\PYZus{}file}\PY{o}{.}\PY{n}{attrs}\PY{p}{[}\PY{l+s}{\PYZsq{}}\PY{l+s}{slide\PYZus{}name}\PY{l+s}{\PYZsq{}}\PY{p}{]} \PY{o}{=} \PY{l+s}{u\PYZsq{}}\PY{l+s}{John Doe}\PY{l+s}{\PYZsq{}}
\PY{n}{root\PYZus{}group} \PY{o}{=} \PY{n}{slide\PYZus{}file}\PY{p}{[}\PY{l+s}{\PYZsq{}}\PY{l+s}{/}\PY{l+s}{\PYZsq{}}\PY{p}{]}
\PY{n}{image\PYZus{}group} \PY{o}{=} \PY{n}{root\PYZus{}group}\PY{o}{.}\PY{n}{create\PYZus{}group}\PY{p}{(}\PY{l+s}{\PYZsq{}}\PY{l+s}{images}\PY{l+s}{\PYZsq{}}\PY{p}{)}
\PY{n}{n} \PY{o}{=} \PY{l+m+mi}{1000}
\PY{n}{image\PYZus{}dataset} \PY{o}{=} \PY{n}{image\PYZus{}group}\PY{o}{.}\PY{n}{create\PYZus{}dataset}\PY{p}{(}
  \PY{l+s}{\PYZsq{}}\PY{l+s}{image1}\PY{l+s}{\PYZsq{}}\PY{p}{,} \PY{n}{shape}\PY{o}{=}\PY{p}{(}\PY{l+m+mi}{100}\PY{o}{*}\PY{n}{n}\PY{p}{,} \PY{l+m+mi}{100}\PY{o}{*}\PY{n}{n}\PY{p}{)}\PY{p}{,} \PY{n}{dtype}\PY{o}{=}\PY{n}{np}\PY{o}{.}\PY{n}{uint16}\PY{p}{)}
\PY{n}{data} \PY{o}{=} \PY{n}{np}\PY{o}{.}\PY{n}{arange}\PY{p}{(}\PY{n}{n}\PY{o}{*}\PY{n}{n}\PY{p}{,} \PY{n}{dtype}\PY{o}{=}\PY{n}{np}\PY{o}{.}\PY{n}{uint16}\PY{p}{)}\PY{o}{.}\PY{n}{reshape}\PY{p}{(}\PY{p}{(}\PY{n}{n}\PY{p}{,}\PY{n}{n}\PY{p}{)}\PY{p}{)}
\PY{n}{image\PYZus{}dataset}\PY{p}{[}\PY{n}{n}\PY{p}{:}\PY{l+m+mi}{2}\PY{o}{*}\PY{n}{n}\PY{p}{,}\PY{n}{n}\PY{p}{:}\PY{l+m+mi}{2}\PY{o}{*}\PY{n}{n}\PY{p}{]} \PY{o}{=} \PY{n}{data}
\end{Verbatim}
As usual when large data sets are involved, the HDF5 library implements a data blocking concept so
called \emph{chunk} which is an application of the divide-and-conquer paradigm. Indeed the data compression
as well as the efficiency of the data transfer requires datasets to be splitted in chunks. This feature
is a cornerstone for many features. It permits to read and write only a subset of the
dataset (a \emph{hyperslab}), providing means for Python to map concepts such view and
broadcasting. Moreover it permits to implement a read-ahead and cache mechanism to speed up the data
transfer from storage to memory.

Another cornerstone of the HDF5 library is the implementation of a modular and powerful data transfer
pipeline shown on figure \DUrole{ref}{hdf5-pipeline} whose aim is to decompress the data from chunks stored
on disk, scatter-gather the data and transform them, for example to apply a scale-offset filter. The
h5py module provides the classic GZIP compression as well its faster counterpart LZF \cite{LZF} and
other compression algorithms can be added easily as plugins.\begin{figure}[bht]\noindent\makebox[\columnwidth][c]{\includegraphics[scale=0.60]{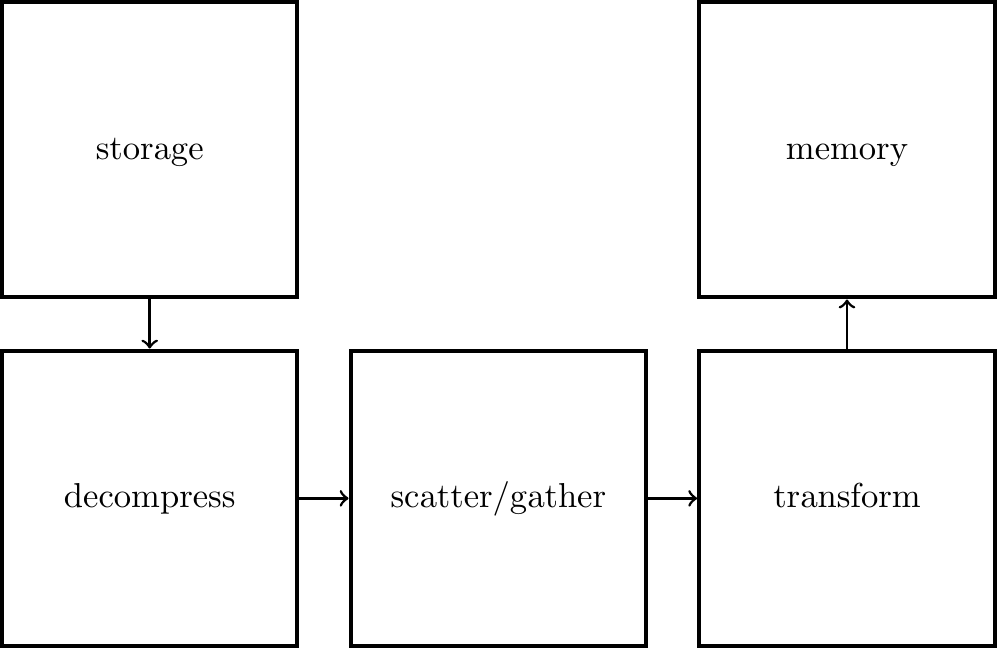}}
\caption{HDF5 Data Transfer Pipeline. \DUrole{label}{hdf5-pipeline}}
\end{figure}

The flexibility of HDF5 permits to use different strategies to store our fields of view according to
our application. The guideline is to think how images will be retrieved and used. For example if we
want to get the fields of view as a planar image then we should use the same shape for the dataset,
i.e. if the image shape is $(H,W)$ then the dataset shape should be $(N_w\,H,W)$ where
$N_w$ is the number of colour planes. Like this we can map directly the data from storage to
memory. The planar format is usually more suited for analysis purpose, but if we want to privilege
the display then we should choose an interleaved format. However we cannot use an interleaved
format in practice if we consider there is an offset between the colour fields of view.

To store the mosaic we can use a dataset per field of view or pack everything in only one
dataset. This second approach would be the natural choice if we had reconstructed the slide image.
For example if the mosaic shape is $(R,C)$ then we can create a dataset of shape
$(R\,N_w\,H,C\,W)$ with a chunk size of $(h,w)$ where $(H, W) = (n\,h, n\,w)$ and
$n \in \mathbb{Z}^{*+}$. Figure \DUrole{ref}{mosaic-dataset} shows an example of a packed mosaic. The
induced overhead will be smoothed by the fact the images are stored on disk as chunks.
% thanks to the data blocking to make this efficient and transparent
\begin{figure}[bht]\noindent\makebox[\columnwidth][c]{\includegraphics[scale=0.50]{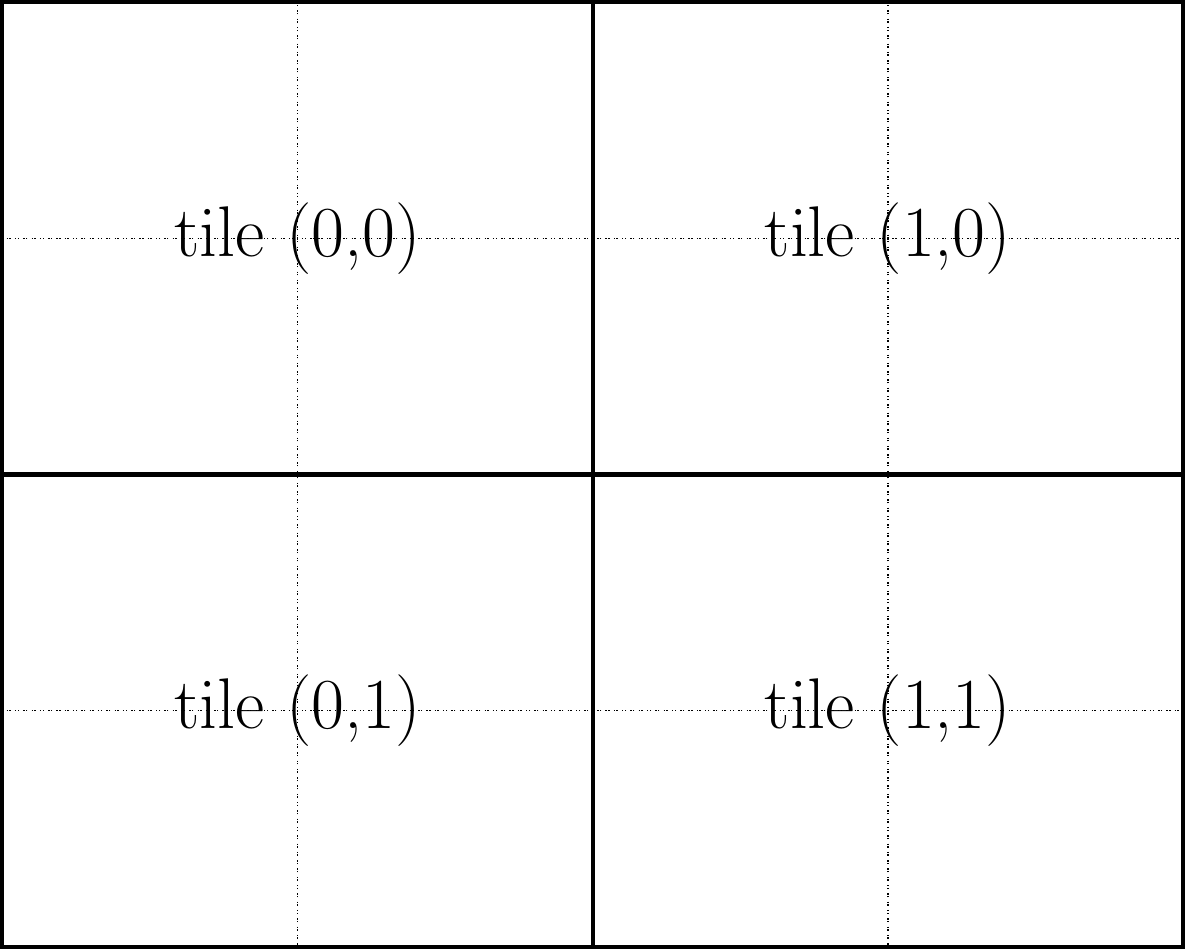}}
\caption{A dataset for a $2 \times 2$ mosaic, chunks are represented by dotted
squares. \DUrole{label}{mosaic-dataset}}
\end{figure}

However if we want to load at the same time a set of consecutive tiles, then we can use this
linear dataset shape $(R\,C\,N_w\,H,W)$ and index the image using the linearised index
$r\,C + c$. Figure \DUrole{ref}{linear-dataset} shows an example of a linearised mosaic. For example
the code to get the fields of view in the slice $[10,20:30]$ would be:\begin{Verbatim}[commandchars=\\\{\},fontsize=\footnotesize]
\PY{n}{lower\PYZus{}index} \PY{o}{=} \PY{l+m+mi}{10}\PY{o}{*}\PY{n}{C} \PY{o}{+} \PY{l+m+mi}{20}
\PY{n}{upper\PYZus{}index} \PY{o}{=} \PY{l+m+mi}{10}\PY{o}{*}\PY{n}{C} \PY{o}{+} \PY{l+m+mi}{30}
\PY{n}{field\PYZus{}of\PYZus{}view\PYZus{}step} \PY{o}{=} \PY{n}{NW} \PY{o}{*} \PY{n}{H}
\PY{n}{lower\PYZus{}r} \PY{o}{=} \PY{n}{lower\PYZus{}index} \PY{o}{*} \PY{n}{field\PYZus{}of\PYZus{}view\PYZus{}step}
\PY{n}{upper\PYZus{}r} \PY{o}{=} \PY{n}{upper\PYZus{}index} \PY{o}{*} \PY{n}{field\PYZus{}of\PYZus{}view\PYZus{}step}
\PY{n}{memory\PYZus{}map} \PY{o}{=} \PY{n}{image\PYZus{}dataset}\PY{p}{[}\PY{n}{lower\PYZus{}r}\PY{p}{:}\PY{n}{upper\PYZus{}r}\PY{p}{,}\PY{p}{:}\PY{p}{]}
\end{Verbatim}
And to get from here the wth colour plane of the ith field of view, the code would be:\begin{Verbatim}[commandchars=\\\{\},fontsize=\footnotesize]
\PY{n}{row\PYZus{}offset} \PY{o}{=} \PY{n}{i} \PY{o}{*} \PY{n}{field\PYZus{}of\PYZus{}view\PYZus{}step} \PY{o}{+} \PY{n}{w} \PY{o}{*} \PY{n}{H}
\PY{n}{colour\PYZus{}image} \PY{o}{=} \PY{n}{memory}\PY{p}{[}\PY{n}{row\PYZus{}offset}\PY{p}{:}\PY{n}{row\PYZus{}offset} \PY{o}{+}\PY{n}{H}\PY{p}{,}\PY{p}{:}\PY{p}{]}
\end{Verbatim}
If the mosaic is sparse we can still pack the mosaic and use a bisection algorithm to perform a binary
search to get the corresponding linear index used for the storage.\begin{figure}[bht]\noindent\makebox[\columnwidth][c]{\includegraphics[scale=0.50]{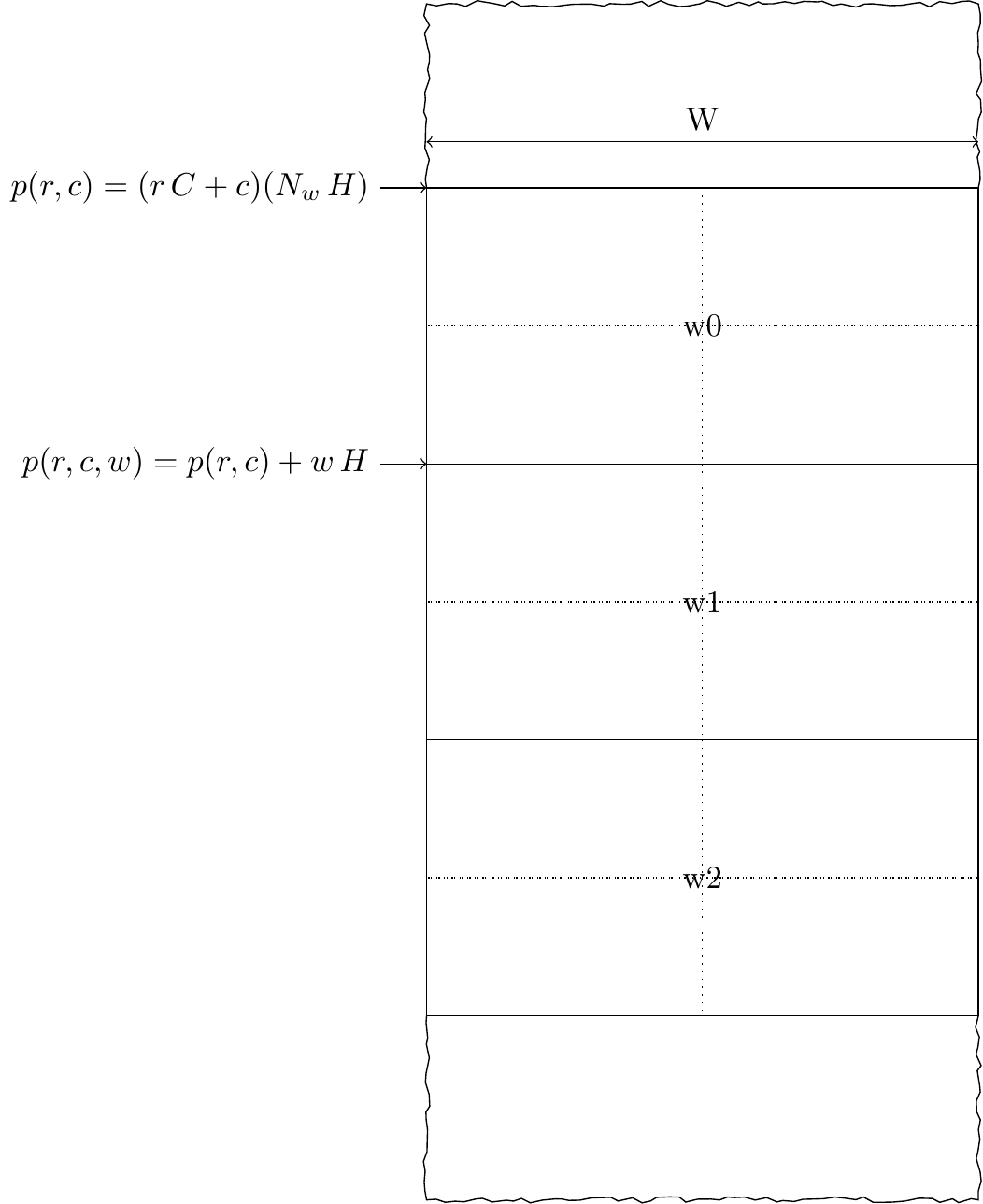}}
\caption{A linear dataset for an acquisition having 3 colours where the pointer to a tile and a plane are
shown. \DUrole{label}{linear-dataset}}
\end{figure}

One can argue this approach is not natural, but encapsulating the slice computation in a virtual
slide API allows for efficient ways to store and retrieve the data. A better approach would be
to have a direct access to the chunks, but actually the HDF5 API does not provide such facility (it
only provides direct chunk write up to now). Thus if we do not want to rewrite or extend the
library, the hyperslab mechanism is a nice alternative. However if we dislike this packing method, we can
still use the following dataset layout $(R,C,N_w,H,W)$ with this chunk layout
$(1,1,1,H,W)$, where the slicing is more natural. Anyway the right approach is to test several
dataset layouts and measure the I/O performance, using the tool \emph{h5perf} provided with the HDF5 SDK.
More details about chunking can be found in the reference \cite{HDF5-Chunking}.

This storage method can be easily extended to a more complicated acquisition scheme having
z-stacks or a time dimension.

\subsection{Remote Virtual Slide%
  \label{remote-virtual-slide}%
}

We have now defined a framework to store our virtual slide based on top of the stack HDF5/h5py that
relies on an HDF5 file stored on a local system or a network file system to work in a client-server
manner. This framework works perfectly, but a network file system has some limitations in comparison
to a real client-server framework. In particular a network file system is complex and has side
effects on an IT infrastructure, for example the need to setup an authentication mechanism for
security. Moreover we cannot build a complex network topology made of a virtual slide broadcast
server and clients.

We will now introduce the concept of remote virtual slide so as to add a real client-server feature
to our framework. We have two types of data to send over the network, the slide header and the
images. Since images are a flow of bytes, it is easy to send them over the network and use the Blosc
\cite{Blosc} real-time compression algorithm to reduce the payload. For the slide header, we can
serialise the set of attributes to a JSON \cite{JSON} string, since the attributes data types are
numbers, strings and tuples of them.

For the networking layer, we use the ZeroMQ \cite{ZMQ} library and its Python binding PyZMQ
\cite{PyZMQ}. ZeroMQ is a socket library that acts as a concurrency framework, carries message across
several types of socket and provides several connection patterns. ZeroMQ is also an elegant solution
to the global interpreter lock \cite{GIL} of the CPython interpreter that prevent real
multi-threading. Indeed the connection patterns and the message queues offer a simple way to
exchange data between processes and synchronise them. This library is notably used by IPython
\cite{IPython} for messaging.

The remote virtual slide framework is build on the request-reply pattern to provide a client-server
model. This pattern can be used to build a complex network topology with data dealer, router and
consumer.

\section{Microscope Interconnection%
  \label{microscope-interconnection}%
}

As a first illustration of the remote virtual slide concept, we will look at the data flow between
the automated microscope so called \emph{scanner} and the software component, so called \emph{slide writer},
whose aim is to write the HDF5 file on the file server. This client-server or producer-consumer framework is
shown on figure \DUrole{ref}{slide-writer-architecture}. To understand the design of this framework, we
have to consider these constrains. The first one is due to the fact that the producer does not run
at the same speed than the consumer. Indeed we want to maximise the scanner throughput and at the
same time maximise the data compression which is a time consuming task. Thus there is a
contradiction in our requirements. Moreover the GIL prevents real time multi-threading. Thus we must
add a FIFO buffer between the producer and the consumer so as to handle the speed difference
between them. This FIFO is called \emph{slide proxy} and acts as an image cache. The second constraint is
due to the fact that the slide writer can complete its job after the end of scan. It means the
slide writer will not be ready to process another slide immediately, which is a drawback if we want
to scan a batch of slides. Thus we need a third process called \emph{slide manager} whose aim is to fork
a slide writer for each scan that will itself fork the slide proxy. Due to this fork mechanism, these
three processes, slide manager, slide writer and slide proxy must run on same host so called \emph{slide
server}. For the other component, all the configurations can be envisaged.

The last component of this framework is the slide database whose aim is to store the path of the
HDF5 file on the slide server so as to retrieve the virtual slide easily.\begin{figure}[bht]\noindent\makebox[\columnwidth][c]{\includegraphics[scale=0.50]{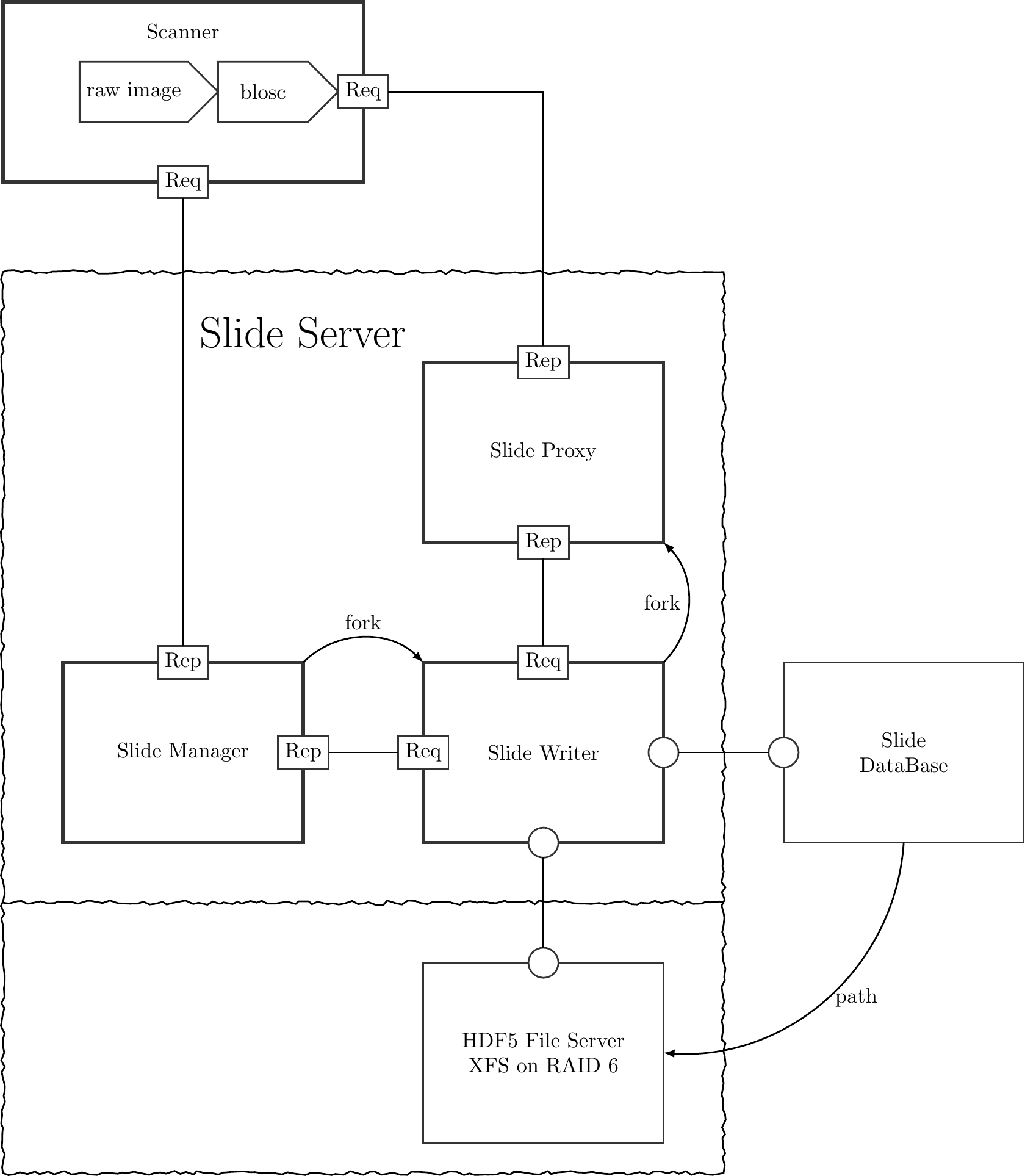}}
\caption{Virtual Slide Writer Architecture. \DUrole{label}{slide-writer-architecture}}
\end{figure}

\section{Slide Viewer Graphic Engine%
  \label{slide-viewer-graphic-engine}%
}

The slide viewer graphic engine works as Google Map using image tiles and follows our concept to
reconstruct the slide image online. We can imagine several strategies to reconstruct the slide
image. The first one would be to perform all the computation on CPU. But nowadays we have GPU that
offer a higher level of parallelism for such a task. GPUs can be programmed using several API like
CUDA, OpenCL and OpenGL \cite{OpenGL}. The first ones are more suited for an exact computation and the
last one for image rendering. In the followings we are talking about modern OpenGL where the fixed
pipeline is deprecated in favour of a programmable pipeline.

The main features of the slide viewer are to manage the viewport, the zoom level and to provide an
image processing to render a patchwork of 16-bit images. All these requirements are fulfilled by
OpenGL. The API provides a way to perform a mapping of a 2D texture to a triangle and by extension
to a quadrilateral which is a particular form of a triangle strip. This feature is perfectly suited
to render a tile patchwork.

The OpenGL programmable pipeline is made of several stages. For our topic, the most important ones
are the vertex shader, the rasterizer and the fragment shader, where a fragment corresponds to a
pixel. The vertex shader is mainly used to map the scene referential to the OpenGL window
viewport. Then the rasterizer generates the fragments of the triangles using a scanline algorithm
and discards fragments which are outside the viewport. Finally a fragment shader provides a way to
perform an image processing and to manage the zoom level using a texture sampler. Figure
\DUrole{ref}{opengl-viewport} shows an illustration of the texture painting on the viewport.\begin{figure}[bht]\noindent\makebox[\columnwidth][c]{\includegraphics[scale=0.50]{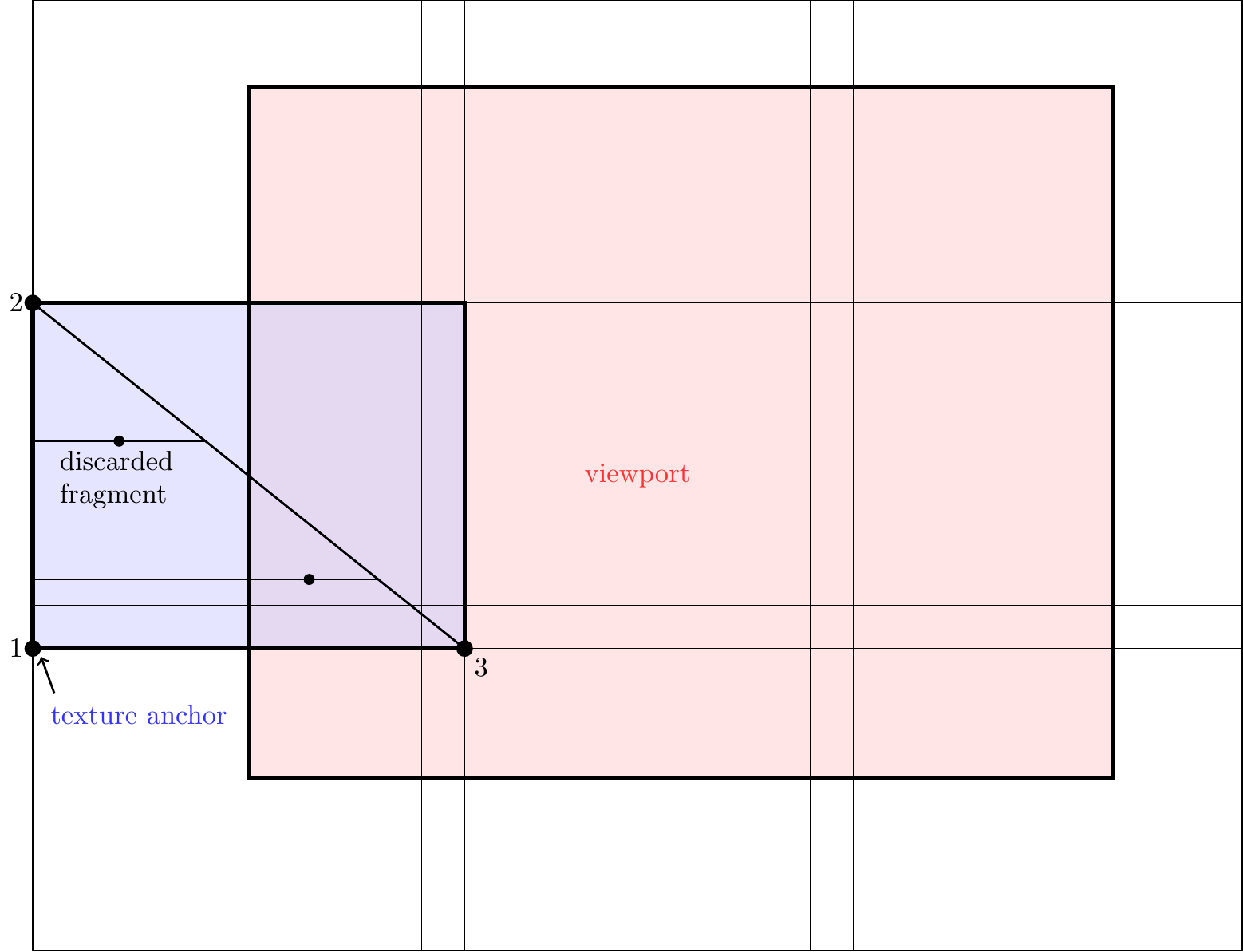}}
\caption{OpenGL viewport and texture painting. The overlapped black rectangles represent the mosaic of
tiles. The red rectangle shows the viewport area. And the blue rectangle illustrates the
rendering of a texture for a tile which is partially out of the viewport area. The horizontal
line represents the sampling of the triangle defined by the vertexes (1, 2, 3) using a scanline
algorithm. Pixels out of the viewport are discarded. \DUrole{label}{opengl-viewport}}
\end{figure}

A texture can have from one to four colour components (RGBA), which make easy to render a slide
acquisition with up to four colours. To render more colours, we just need more than one texture by
tile and a more complicated fragment shader. If the tiles are stored in a planar format then we have
to convert them to an interleaved format, we call this task texture preparation. However we can also
use a texture per colour but in this case we have to take care to the maximal number of texture
slots provided by the OpenGL implementation, else we have to perform a framebuffer blending. The
main advantage of using a multi-colour texture is for efficiency since the colour processing is
vectorised in the fragment shader. However if we want to register the colour on-line, then the
texture lookup is any more efficient.

To render the viewport, the slide viewer must perform several tasks. First it must find the list of
tiles that compose the viewport and load these tiles from the HDF5 file. Then it must prepare the
data for the corresponding textures and load them to OpenGL. The time consuming tasks are the last
three ones. In order to accelerate the rendering, it would be judicious to perform these tasks in
parallel, which is not simple using Python.

For the tile loading, we can build on our remote virtual slide framework in order to perform an
intelligent read-ahead and to eventually prepare the data for the texture.

The parallelisation of the texture loading is the most difficult part and it relies of the OpenGL
implementation. Modern OpenGL Extension to the X Window server (GLX) supports texture loading within
a thread, but this approach cannot be used efficiently in Python due to the GIL. Moreover we
cannot use a separate process to do that since it requires processes could share an OpenGL context,
which is only available for indirect rendering (glXImportContextExt). Also we cannot be sure the
multi-threading would be efficient in our case due to the fact we are rendering a subset of the
mosaic at a time and thus textures have a short life time. And the added complexity could prove to
be a drawback.

Since our mosaic can be irregular, we cannot found by a simple computation which tiles are in the
viewport. Instead we use an R-tree for this purpose. All the tiles boundaries are filled in the
R-tree. And to get the list of tiles within the viewport, we perform an intersection query of the
R-tree with the viewport boundary.

\subsection{Slide Viewer Architecture%
  \label{slide-viewer-architecture}%
}
\begin{figure}[bht]\noindent\makebox[\columnwidth][c]{\includegraphics[scale=0.50]{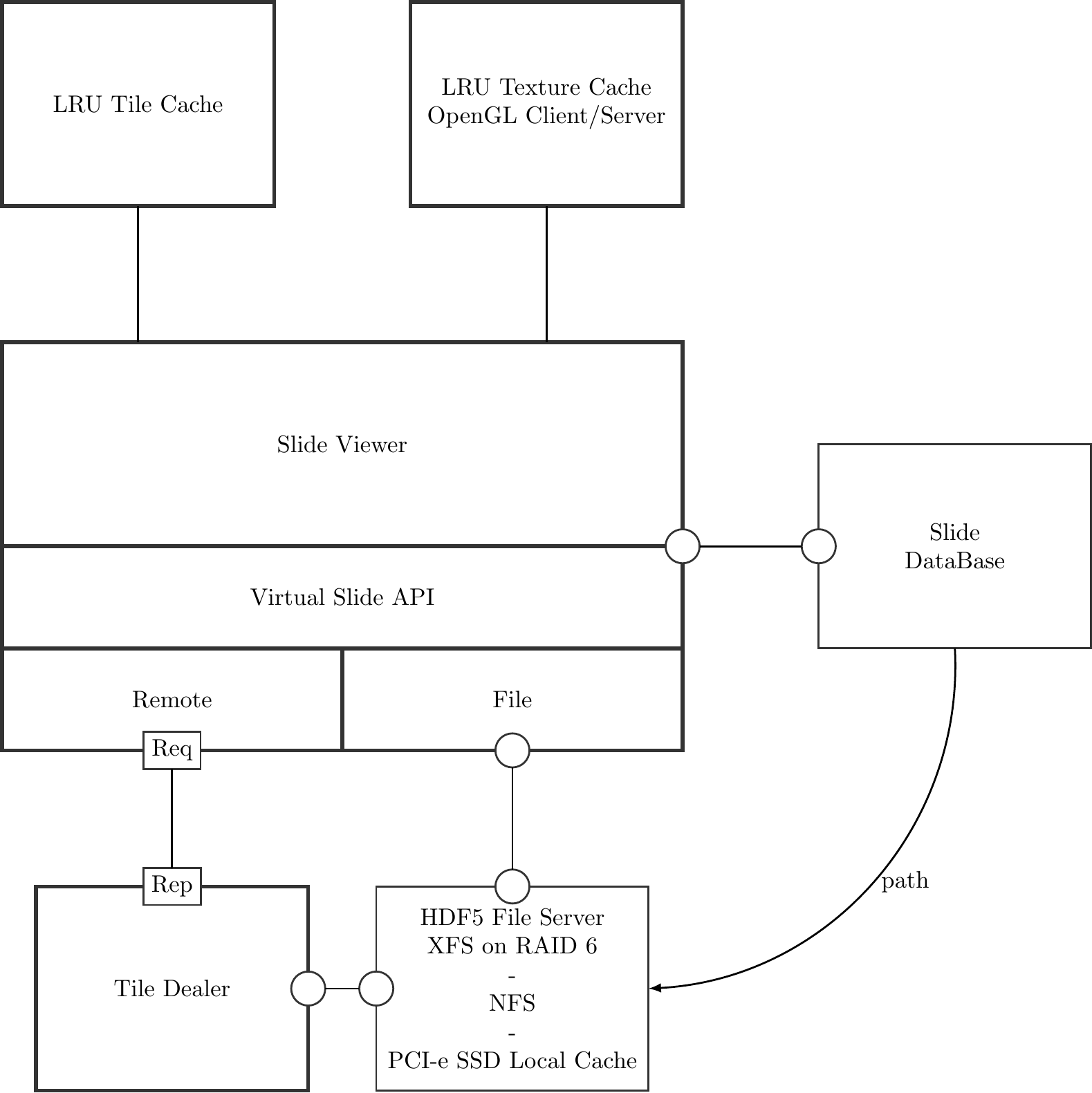}}
\caption{Slide Viewer Architecture. \DUrole{label}{slide-viewer-architecture}}
\end{figure}

Figure \DUrole{ref}{slide-viewer-architecture} shows the architecture of our slide viewer where the virtual
slide API can access the data through the HDF5 file or the remote framework. In our IT infrastructure, HDF5
files are stored on a file server that can provide a network file system to access files
remotely. The remote virtual slide can be used in two different ways according to the machine where
the process of the server side, called \emph{tile dealer}, is executed. If this process runs on the
same host as the slide viewer, then we can use it to implement a read-ahead mechanism to
parallelise the tile loading. And if it runs on the file server, then we can use it as an
alternative to the network file system in a similar way as a virtual slide broadcast service. This
second example demonstrates the remote virtual slide is a fundamental software component in our
framework that open the way to many things.

Another way to access efficiently the data, it to use a local cache to store temporally the virtual
slide. Nowadays we can build on a very fast locale cache using a PCI-e SSD card, which commonly
reach a read/write bandwidth of $1000\,\text{MB/s}$ and thus outperforms most of the hardware
RAID.

The slide viewer implements two Least Recently Used caches to store the tiles and the
textures. These caches are a cornerstone for the fluidity of the navigation within the slide, since
it helps to reduce the viewer latency. Nowadays we can have on a workstation
$64\,\text{GB}$ of RAM for a decent cost, which open the way to a large in memory cache in
complement to a PCI-e SSD cache. In this way we can build a 3-tier system made of a file server to
store tera bytes of data, a PCI-e SSD cache to store temporally slides and an in memory cache to
store a subset of the virtual slide.

\subsection{Vertex and Fragment Shader%
  \label{vertex-and-fragment-shader}%
}

In modern OpenGL all the computations must be performed by hand from the viewport modelling to the
fragment processing, excepted the texture sampling which is provided by the OpenGL Shading Language.

Since we are doing a two dimensional rendering, it simplifies considerably the viewport model and
the coordinate transformation. OpenGL discards all the fragment that are outside the
$[-1,1]\times[-1,1]$ interval. Thus to manage the viewport, we have to transform the slide
frame coordinate using the following model matrix:\begin{equation}
\label{viewport matrix}
\left(\begin{array}{c}
x \\
y \\
z \\
w \\
\end{array}\right)
=
\left(\begin{array}{cccc}
\frac{2}{x_{sup} - x_{inf}} & 0 & 0 & -\frac{x_{inf} + x_{sup}}{x_{sup} - x_{inf}} \\
0 & \frac{2}{y_{sup} - y_{inf}} & 0 & -\frac{y_{inf} + y_{sup}}{y_{sup} - y_{inf}} \\
0 & 0 & 1 & 0 \\
0 & 0 & 0 & 1 \\
\end{array}\right)
\left(\begin{array}{c}
x_s \\
y_s \\
0 \\
1 \\
\end{array}\right)
\end{equation}where $[x_{inf},x_{sup}]\times[y_{inf},y_{sup}]$ is the viewport interval and
$(x_s,y_s)$ is a coordinate in the slide frame.

OpenGL represents fragment colour by a normalised float in the range $[0,1]$ and values which
are outside this range are clamped. Thus to transform our 16-bit pixel intensity we have to use this
formula:\begin{equation}
\label{normalised luminance}
% _\text{normalised
\hat{l} = \frac{l - I_{inf}}{I_{sup} - I_{inf}}
\end{equation}where $0 \leq I_{inf} < I_{sup} < 2^{16}$. This normalisation can be used to perform an image
contrast by adjusting the values of $I_{inf}$ and $I_{sup}$.

The fact OpenGL supports the unsigned 16-bit data type for texture permits to load the raw data
directly in the fragment shader without information loss. According to the configuration of OpenGL,
the RAMDAC of the video adapter will convert the normalised floats to an unsigned 8-bit intensity
for a standard monitor or to 10-bit for high-end monitor like DICOM compliant models.

As soon as we have converted our pixel intensities to float, we can apply some image processing
treatments like a gamma correction for example.

In the previous paragraphs, we told we can load in a texture up to four colour components using
RGBA textures. Since monitors can only render three colour components (RGB), we have to transform a
four components colour space to a three components colour space using a \emph{mixer matrix}. This
computation can be easily extended to any number of colours using more than one texture. The mixer
matrix coefficients should be choose so as to respect the normalised float range.

Another important feature of the slide viewer is to permit to the user to select which colours will
be displayed on the screen. This feature is easily implemented using a diagonal matrix so called
\emph{status matrix} with its coefficients set to zero or one depending of the colour status.

We can now write the matrix computation for the rendering of up to four colours:\begin{equation}
\label{texture fragment shader}
\left(\begin{array}{c}
r \\
g \\
b \\
\end{array}\right)
=
\underbrace{
\left(\begin{array}{ccc}
m_{r0} & \ldots & m_{r3} \\
m_{g0} & \ldots & m_{g3} \\
m_{b0} & \ldots & m_{b3} \\
\end{array}\right)
}_\text{mixer matrix}
\underbrace{
\left(\begin{array}{ccc}
s_0 & & \\
& \ddots & \\
& & s_3 \\
\end{array}\right)
}_\text{status matrix}
\left(\begin{array}{c}
\hat{l}_0 \\
\vdots \\
\hat{l}_3 \\
\end{array}\right)
\end{equation}If we consider a GPU with more than 1024 cores, then most of the rows of our display will be
processed in parallel which is nowadays impossible to perform with a multi-core CPU. It is why our
approach to render a mosaic of tiles is so efficient and the rendering is nearly done in real time.

\subsection{Zoom Layer%
  \label{zoom-layer}%
}

When the texture must be magnified, it is important to enlarge the pixel without interpolation. In
OpenGL it is achieved by using the \emph{GL\_NEAREST} mode for the texture magnification filter.

Despite GPU are very powerful, there is a maximal number of tiles in the viewport that can be
reasonably processed. The amount of memory of the GPU is an indicator of this limitation. If we
consider a GPU with $2048\,\text{MB}$, then we can load 66 textures having a layout of $2560
\times 2160\,\text{px}$ and a 16-bit RGB format. It means we can display a mosaic of $8 \times
8$ at the same time. If we want to display more tiles at the same time, then we have to compute a so
called \emph{mipmaps} which is a pyramidal collection of mignified textures. Usually we perform a
geometric series that corresponds to divide by two the size of the texture recursively. Due to the
power of the GPU, it is not necessary to compute the entire pyramid, but just some levels. In our
case we can compute the levels 8 and 16. For higher levels according to the size of the mosaic, it
could be more efficient to compute a reconstructed image. These mignified textures can be computed
online using CUDA or stored in the HDF5 files.

Our slide viewer implements a zoom manager in order to control according to the current zoom which
zoom layer is active and to limit the zoom amplitude to an appropriate range. Moreover we can
implement some excluded zoom ranges and force the zoom to the nearest authorised zoom according to
the zoom direction.\begin{figure}[t]\noindent\makebox[\columnwidth][c]{\includegraphics[scale=0.18]{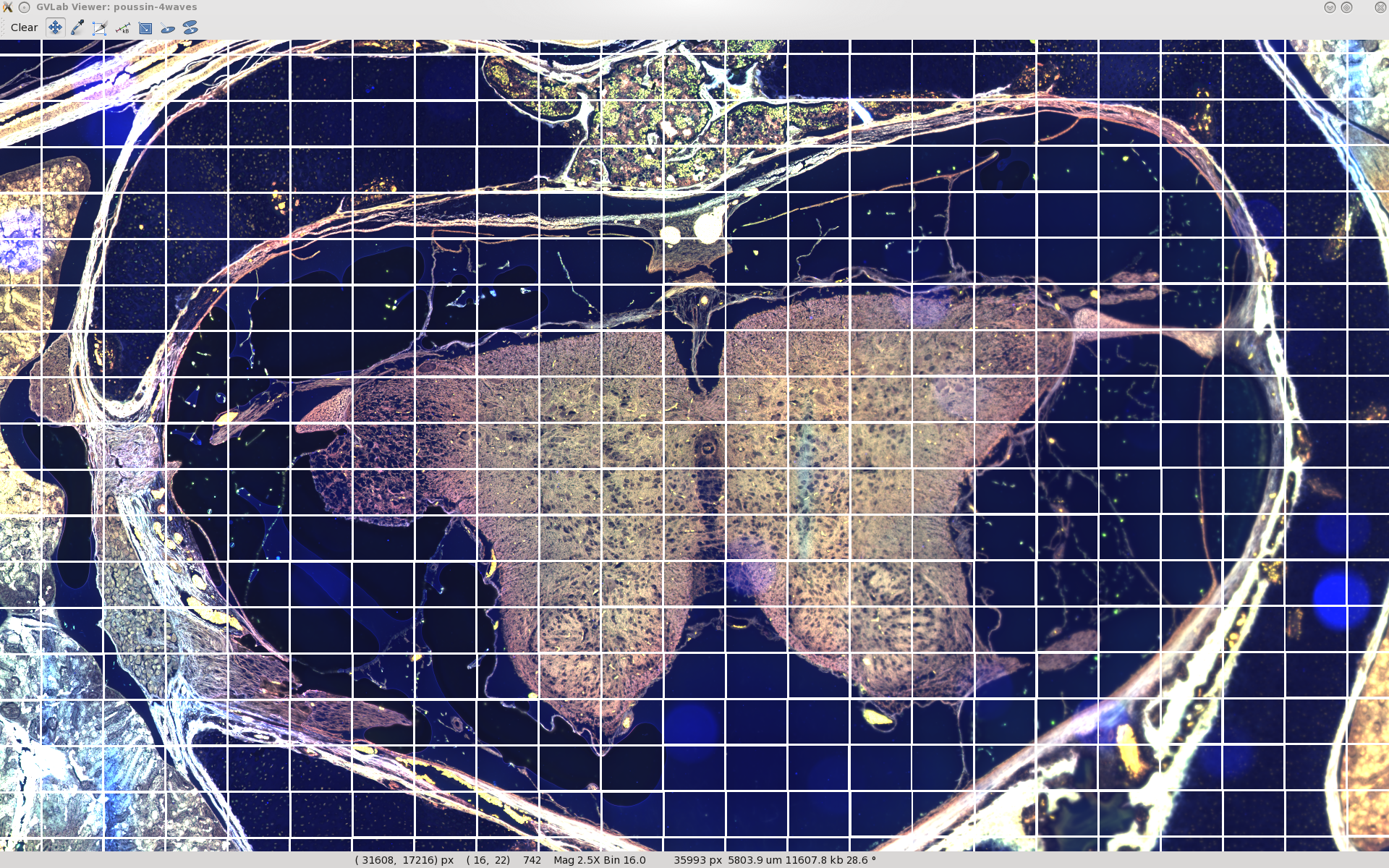}}
\caption{Cell displayed in the slide viewer. The slide was acquired with an epifluorescence-microscope at
magnification $40\times$ with a camera of resolution $1392 \times 1040\,\text{px}$
and with four colours. The size of the part of the mosaic shown on the viewport is $19
\times 22$ \DUrole{label}{slide-viewer-image}, which corresponds to 418 tiles and thus around
$595\,\text{Mpx}$. The dimension of the visible surface is around $4.9 \times
3.1\,\text{mm}$. Here the slide image is rendered at magnification $2.5\times$ and the zoom
layer corresponds to a mignification of level $2^4 = 16$ and thus to a texture of dimension
$87 \times 65\,\text{px}$. So there is around $2\,\text{Mpx}$ to
process. \DUrole{label}{slide-viewer-image}}
\end{figure}
% 820\,544\,

% 2227940

\subsection{Detection Layer%
  \label{detection-layer}%
}

Our slide viewer is not limited to display raw images, but can also display tiles from an image
processing pipeline. When the viewer render a viewport, it first looks which tiles compose the
viewport, then for each tile, it looks if the OpenGL LRU cache has a texture for the corresponding
tile and image processing pipeline, if the texture does not exists yet then it cascades the request
to the tile LRU cache and finally it will asks the image processing pipeline to generate the
image. The tile loading from the virtual slide corresponds to the so called raw image pipeline and
each zoom layer owns its image pipeline. Moreover each pipeline can have its own fragment shader
to customise the rendering.

\subsection{Benchmark%
  \label{benchmark}%
}

Figure \DUrole{ref}{slide-viewer-image} show a reconstructed image made of 418 tiles. For a tile dimension
of $1392 \times 1040\,\text{px}$ and a four colours acquisition, our slide viewer needs around
$2\,\text{s}$ to render the zoom layer 16 and $6\,\text{s}$ for the layer 8 (100 raw
tiles) on a workstation with a Xeon E5-1620 CPU, a GeForce GTX-660 GPU and the HDF5 file stored on a
local SATA hard disk. The required time to load a tile form the HDF5 file is around
$50\,\text{ms}$, thus the tile loading account for $80\,\%$ of the full rendering time.

\section{Conclusion%
  \label{conclusion}%
}

This paper gives an overview how the Python ecosystem can be used to build a software platform for
high-content digital microscopy. Our achievement demonstrates Python is well suited to build a
framework for big data. Despite Python is a high level language, we can handle a large amount of
data efficiently by using powerful C libraries and GPU processing.

First we gave an overview how to store and handle virtual slides using Python, Numpy and the HDF5
library. Different methods to store the images of the fields of view within a dataset was
discussed. In particular the case where we do not reconstruct an image of slide once and for all,
but rather perform an on-line reconstruction from the raw images. Despite our method to store the
images works well, it would be interesting to look deeper in the HDF5 library to see if we could do
something still better.

We described the concept of remote virtual slide which is a client-server model build on top of our
virtual slide framework. We gave two examples of utilisation of this client-server model, the
scanner interconnection with the slide writer and the tile dealer. Also we shown how this
architecture solve the GIL problem and enhance the performance.

Finally we described our slide viewer architecture based on the OpenGL programmable pipeline and a
texture patchwork rendering. We gave an overview on the vertex and the fragment shader. Thanks to the power of
GPU, this method can render more than three colours in quasi real time. Moreover we explained how to
manage the zoom level efficiently so as to overcome the limited amount of RAM of the GPU.

In a near future, it would be interesting to see how the JIT Python interpreter PyPy will enhance
the performance of this framework. Up to now the lake of support of C library like Numpy and Qt
prevents to run the code with it.

The Git repository \url{https://github.com/FabriceSalvaire/PyOpenGLV4} provides an oriented object API on
top of PyOpenGL to work with the OpenGL programmable pipeline. This module is used in our slide
viewer.
% -------------------------------------------------------------------------------------------------

% -------------------------------------------------------------------------------------------------
% End

\end{document}